# Sequential mediation of parasocial relationships for purchase intention: PLS-SEM and machine learning approach

Nora Sharkasi*    and    Saeid Rezakhah§

*Institute for International Strategy (IIS), Tokyo International University,

1 Chome131 Matobakita, Saitama, 3501197, +81(0)492321111, Kawagoe, Japan. Email:
nsharkas@tiu.ac.jp

§Faculty of Mathematics and Computer Science, Amirkabir University of Technology,

424 Hafez Avenue, 15875, Tehran, Iran. Email: Rezakhah@aut.ac.ir

## Abstract

Companies employ social media influencers (SMIs) due to the compelling evidence of their advertising effectiveness; however, more research is required to identify and compare factors driving their success. We investigate the effect of source influence (SI) on the intention to purchase (I2P) through the sequential mediation of parasocial relationships (PSR) with benign envy (BE) and PSR with perceived brand–influencer fit (BIF). Two independent samples (N=411; N=355 from Europe and Southeast Asia) are used to perform: (i) PLS–SEM analysis to obtain the model's predictive power and (ii) classification–based machine learning (ML) to evaluate the model's accuracy. Moreover, within–study and between–studies comparative analyses are performed. We use regression analysis and split–test ML technique for validation. Both samples indicate a higher role of trustworthiness and expertise in forming SI. Furthermore, comparative mediation analysis and predictive accuracy scores show that the audience–related feature, BE, played a more vital role in affecting/predicting the followers' I2P than the brand-related feature, BIF. Our findings contribute to the knowledge of SMIs' credibility and comparative analysis paradigms and provide a better understanding for marketing practitioners and researchers.

# Introduction

In marketing practice, brands collaborate with social media influencers (SMIs) to endorse their products and services due to their ability to stimulate positive responses (Kim, 2021). They are also perceived as more credible than conventional TV celebrities and thus more impactful and persuasive (Yılmazdoğan et al., 2021; Yuan and Lou, 2020); this is due to SMIs' homophily with followers (Sokolova and Kefi, 2020), transparency and expertise of the influencers (Breves et al., 2021; Xie and Feng, 2020; Tian and Li, 2022; Schouten et al., 2019).

Over the years, influencer marketing has gained much recognition from marketing professionals and researchers (Hudders et al., 2021). As a result, companies has been employing professional SMIs with a relatively large number of followers to endorse brands. However, as this strategy fails in certain situations, researchers suggest other strategies to deliver optimal results, such as employing micro-influencers (Kay et al., 2020), among other strategies (Drummond et al., 2020). Therefore, it is crucial to study the factors driving the success of SMIs in order to devise better online advertising campaigns and strategies. We consider the elaboration likelihood model (ELM) (Petty and Cacioppo, 1986) to forge a clear direction to identify essential factors in the influencer-follower relationship. A parasocial relationship (PSR) is one-sided between SMIs and their many followers (Lou and Kim, 2019; Stever, 2017).

Although the influencer–follower relationship is mediated by technology, it still allows for enough exposure to the influencer's content over multiple sessions of media viewing; this establishes a sense of connectedness and develops a tie with the SMI, measured by PSR (Bond, 2018, p. 459; Dibble et al., 2016, p. 21; Horton and Wohl, 1956). Several research studies have shown that SI affects the followers' intention to purchase (I2P) through the mediation effect of PSR (Masuda et al., 2021; Ong et al., 2021; Reinikainen et al., 2020). Based on the ELM, this study considers the following main factors for analysis:

(i)     The source of the persuasion or influence, as in the source influence (SI) of SMIs

(ii)    The communicated message as in the brand-influencer fit (BIF) of the message

(iii)   The audience–related features as the followers' benign envy (BE) toward an SMI

The literature review section includes detailed explanations of the above three factors. We chose to study the variables BIF and BE as recent studies explore these constructs within the credibility theory framework. This theory is widely used with the PSR to explore the influencer-follower relationship. We refer to the following relevant studies that explored the credibility theory with BIF (e.g., Li et al., 2022; Bi and Zhang, 2022; Martínez-Lopez et al., 2020; Breves et al., 2019) and BE (e.g., Bi and Zhang, 2022; Coelhoso et al., 2022; Asdecker, 2022; Lee and Eastin, 2020). Moreover, the followers' willingness to acquire the influencer's lifestyle and the material items they endorse emerge from feeling BE toward the influencer (Parrott and Smith, 1993). When the SMI is an excellent fit for the brand - BIF, s/he is often viewed as more honest and is perceived as more credible to endorse the brand (Breves et al., 2019). Therefore, a high BIF enforces the positive effect of influencer credibility, which positively affects BE (Tran et al., 2022).

systematic literature review on influencer marketing conducted by Vrontis et al. (2021) suggests conducting more research on the audience–related features. Several recent studies (e.g., Janssen et al., 2022; Mettenheim and Wiedmann, 2021; Song and Kim, 2020; Lee and Eastin, 2020) shed light on influencer–related characteristics through examining their SI in congruence with the endorsed brand. However, a few studies examined relevant sequential mediation effect of PSR with brand–related factors (e.g., Hugh et al., 2022; Bi and Zhang, 2022; Shen, 2020). As far as we know, no study specifically examined the sequential mediation effect of PSR with BIF, and no research has considered the sequential mediation of BE via PSR. We refer to some relevant research considering the sequential mediation effect of 'tie strength' with benign envy (e.g., Wang et al., 2021; Duan, 2021; Li, 2019; Wu and Srite, 2021).

This study focuses on the effect of the sequential mediation of BIF with PSR on the association between SI and I2P and of BE with PSR on the same association. In addition, we address the gap that there is a scarcity of studies comparing between audience and brand related features. We investigate whether a human–related factor, like BE, or a product–related factor, like BIF, plays a more vital role in affecting and predicting the effect of SI on I2P. To answer this question empirically, we collect and analyze data from two independent studies, with samples N=411 and N=355 from Europe and Southeast Asia. In study– I, a section of the questionnaire collects data to measure BIF, and in study – II, we administer the same questionnaire measuring BE instead of BIF. We use the partial-least squares structural equation modeling (PLS–SEM) method for analysis.

PLS-SEM is one of the most commonly used methods for mediation analysis in social science research, including marketing (Shmueli et al., 2019). It tests prediction hypotheses based on in-sample metrics that indicate how well the hypothesized model fits the data. Even though in–sample fit indices have been used as indicative of the model's predictive power (Hair and Sarstedt, 2021), the managerial and practical implications of the model should be based on out–of–sample model fit indices (Sarstedt and Danks, 2022). With out–of–sample fit indices, the generalization of practical and managerial implications will be statistically indisputable and could be generalized across different samples, contexts, and time (Sarstedt and Danks, 2022). For this purpose, the open–source package PLSpredict (Shmueli et al., 2016), which uses out–of–sample evaluation of the model's predictive power, has become a standard tool in the PLS–SEM analysis. Nonetheless, the PLSpredict does not quantify the model's predictive performance. By applying machine learning (ML) classification algorithms, we obtain quantifiable values of the model's predictive accuracy, which helps in conducting comparisons, prioritizing variables, and verifying its suitability to drive managerial and practical implications.

A limited number of management and marketing studies applied or discussed machine learning approaches relevant to structural equation modeling (SEM). We group (SEM) – machine learning blended studies in three groups. First, studies just discussed applying machine learning for prediction as a supplementary practice in (SEM). For example, Sarstedt and Danks (2022) discussed the necessity of using out-of-sample fit indices to produce managerial implications in Human Resource literature. Richter et al. (2022) lightly mentioned machine learning in their paper's section, "Triangulating PLS-SEM with Other Methods/Techniques." Hair and Sarstedt (2021) discussed causal inferences in machine learning for marketing. The second group consists of studies that applied different machine-learning approaches to management and marketing-related areas. For example, Arshi et al. (2021) used the train-test split approach to predict the effect of the independent variables on entrepreneurial behavior with only one dataset for validation. Zobair et al. (2021) evaluated their questionnaire empirically by validating their proposed research model and hypotheses by using a two-staged PLS-SEM and deep neural network ML. Their approach detects linear and non-linear relationships related to normality, linearity, and homoscedasticity. Elnagar et al. (2022) predicted the intention of customers to use a smartwatch by applying ML algorithms using Weka. The third group encompasses studies that focused on developing technical extensions of (SEM) based on machine learning techniques. For example, van Kesteren and Oberski (2021) introduced three (SEM) extensions. Our study belongs to the second group of research. We employ six classification algorithms to validate whether latent predictive variables in the PLS-SEM predict the latent response variable with an acceptable accuracy score. This study presents the following new contributions:

(1) Considering the sequential mediation effect of BE with PSR and BIF with PSR on the association of SI and I2P.

(2) propose a comparative analysis paradigm by performing within–study and between–studies comparisons. Within–study analysis aims to compare the mediation effect of just PSR on the association of SI and I2P vs. the sequential mediation of both: PSR with BIF and PSR with BE in study – I and – II, respectively. Between–studies compare the sequential mediation effects of PSR with BIF in study – I vs. PSR with BE in study – II. In addition to comparing predictive accuracy, we consider other metrics resulting from the measurement and structural models; such as: the total variance explained, AIC, and PLSpredict resulting errors.

(3) This study considers a multidisciplinary analysis approach, incorporating PLS–SEM method, widely used in social science research, with classification–based ML algorithms. With ML classification algorithms, we attempt to quantify the model's predictive performance by assessing its predictive accuracy (Gkikas et al., 2022;

Cacciarelli and Boresta, 2022; Sharkasi et al., 2015; Govindarajan and Chandrasekaran, 2010).

(4) Cross-validating the results by conducting split–test ML technique to compare the predictive performance by considering the data of study – I as a learning dataset and the data of study – II as a testing dataset, and vice versa. This step is important to eliminate any speculations of possible sample bias that steers results in favor of a particular variable.

This paper is organized as follows; the next section is a literature review to synthesize source influence and develop the main hypotheses. Section 3 explains the methodology. Section 4 focuses on study – I; it covers the instrument and sampling, the measurement and structural models. It also covers the prediction accuracy of each model. Section 5 concerns study – II, which is similarly organized as study – I. Section 6 demonstrates comparative analysis and results validation. Finally, the paper is concluded in Section 7, which consists of key findings, theoretical and managerial implications, limitations, and future works.

# 1. Literature Review and Theoretical Framework

Influencer marketing encompasses the endorsement and sponsorship or advertisement of an offering by an influential person or an opinion leader (Breves et al., 2019). It also involves the marketing dynamics, including the strategy and tactics necessary to select the appropriate influencers who demonstrate a proper fit with the brand (Lou and Yuan, 2019). However, influence does not always accompany a high number of followers; nowadays, marketing managers seek the appropriate fit between their target audience and the influencer (Janssen et al., 2022; Qian and Park, 2021), due to the increasing power of micro-influencers compared to those with a considerable following (Ritchie, 2023). Thus, the anticipated quality of the outcome, and not necessarily the quantity of followers matters the most (Janssen et al., 2022). Nowadays, social media influencers (SMI) form a crucial power in advertising (Kim, 2021). Influencer marketing has not only gained much recognition by becoming an important part of digital marketing strategies, but it has also gained significant research interest in recent years (Hudders et al., 2021). It is thus important to further investigate the factors driving the success of SMIs for better SMIs' selection strategies and also for more effective audience targeting strategies.

## 1.1. Source Influence (SI)

Despite the growing role of social media influencers in promoting brands, there is still a need to address the effectiveness of influencers endorsing brands on the perception of their followers (Jiménez-Castillo and Sánchez-Fernández, 2019). The effectiveness of brand endorsement by social media influencers (SMIs) is understood through the source credibility theory (Li et al., 2022; Bi and Zhang, 2022; Martínez-Lopez et al., 2020; Breves et al., 2019; Munnukka et al., 2016), whereby source influence (SI) is conceptualized by four main factors: expertise (EXP), trustworthiness (TW), attractiveness (ATT), and similarity (SIM) (Ohanian, 1990).

The classical theory of credibility utilizes the concept of source credibility (Hovland et al., 1953; Belknap, 1954) to explain the persuasion ability of celebrities and SMI. Influencers' credibility is a crucial concept in influencer marketing (Belanche et al., 2021). Due to SMIs homophily with followers (Sokolova and Kefi, 2020), transparency (Breves et al, 2021; Xie and Feng, 2020; Hwang and Jeong, 2016), and expertise in a specific area (Tian and Li, 2022; Schouten et al, 2019; De Veirman et al., 2016), SMIs are perceived more credible than conventional TV celebrities and thus more impactful and persuasive (Yılmazdoğan et al., 2021; Yuan and Lou, 2020). Brands collaborate with SMIs to endorse their products and services because of their ability to stimulate positive responses (Kim, 2021).

Expertise of the SI implies the knowledge, skills, opinions, views, and claims made around a specific subject (McCroskey, 1966). EXP is the extent to which the source provides valid assertions (Munnukka et al., 2016). It also includes obtaining the necessary skills, adequate knowledge, and experience to endorse a certain brand or product (Van der Waldt et al., 2009). EXP is the perception that the communicator is a qualified expert (Ohanian, 1990). High levels of expertise and similarity with SMI were found to have a vital role in soliciting positive purchase intentions of followers (Lou and Yuan, 2019).

The unique selling point of social media influencers is their relatability, authenticity, and sincerity, as they are perceived as persuasive when they are perceived as trustworthy (Breves et al., 2019). TW of a SI is its sincerity, integrity, and believability; these characteristics are part of the selection criteria of influencers. SMI need to be perceived by their followers as transparent and thus more persuasive (Breves et al., 2021) and also likable to win the trust of the message recipients (Friedman and Friedman, 1978). Moreover, it is also important for SMIs to be perceived as experts in their area to avoid harming their perceived credibility and TW (Sokolova and Kefi, 2020; Martínez-Lopez et al., 2020). SMIs need elevated levels of EXP and TW to develop effective perceptions of credibility among their followers (Schouten et al., 2020), and induce positive purchase intentions of the brand (Herrando and Martín-De Hoyos, 2022; AlFarraj et al., 2021).

Classical research findings suggest that physical attractiveness strengthens the persuasion power of the communicator (Baker and Churchill, 1977). It also positively affects the intention to purchase (I2P) of his/her audience (Petty and Cacioppo, 1980) more than their less attractive counterparts (Kahle and Homer, 1985). This happens through a process called identification. Cohen and Golden (1972) explain that the recipients of a communication message will form a desire to identify with the source of information when they perceive it as attractive and thus want to identify with such a source. Ohanian (1990) constructed a three–component credibility scale for endorsing celebrities; in his scale, he did not only study physical attractiveness (beautiful/Ugly, Sexy/not Sexy), but he also focused on other factors (elegant, classy, and attractive). Erdogan (1999) shed light on the shift of the definition of perceived attractiveness to incorporate any virtuous attributes of the endorser: either visible as the SMI's athletic body or invisible as the sense of humor.

The persuasion power of a message also hangs upon the level of similarity with the SMI (McGuire, 1985). Similarity is defined as "a supposed resemblance between the source and the receiver of the message, familiarity as knowledge of the source through exposure, and likability as affection for the source as a result of the source's physical appearance and behavior" (Erdogan, 1999). Perceived similarity and attractiveness significantly correlate to purchase intentions (Lou and Kim, 2019).

Table 1 lists the questionnaire's observed items used to measure SI (through EXP, TW, ATT, and SIM) based mainly on the measurement scale introduced by Ohanian (1990). The overall SI of the influencer has a positive influence on the (I2P) of followers. Researchers in the area of influencer marketing suggests that the more credible a source is, the more persuasive its message is likely to be (Bi and Zhang, 2022; Breves et al., 2019) and more likely to generate intentions to buy the brand (Breves et al., 2021; Martínez-Lopez et al., 2020; Ohanian, 1991). Based on the previous discussion, we propose the following hypothesis:

*H1: Source influence of social media influencers has a positive effect on the intention to purchase of followers.*

## 1.2.    Parasocial Relationship (PSR)

A conventional definition of Parasocial relationships (PSR) in communication refers to PSR as an illusional feeling of friendship developed by a mass audience toward a media persona perceived as a comforting supporter (Horton and Wohl, 1956). In the area of influencer marketing, PSR could be described by three main characteristics, first, in terms of being related to feeling a sense of friendship and liking between the SMIs and their followers (Hudders et al., 2021). However, it is more of an illusional feeling of intimacy (Lee and Watkins, 2016; Labrecque, 2014). Despite the vicarious interaction, the relationship between the SMIs and their followers seems to be actual, and followers seem to understand the SMI as they understand their real friends (Dibble et al., 2016). As in friendship, PSR is voluntary and involves a personal focus (Wright, 1978), but with "intimacy at a distance" (Horton and Wohl, 1956, p.215).

Second, regarding the time and medium of exposure, PSR refers to the relationship that forms between the follower and the SMI through technology–mediated exposure beyond just one session or episode of media viewing (Dibble et al., 2016, p. 21). As in social relations, PSR develops over a period of time and is enhanced when communication resembles interpersonal interaction (Horton and Wohl, 1956). PSR is a normal consequence of the SMIs' content viewing over a period of time. Thus, followers establish a sense of connectedness with the SMI (Bond, 2018, p. 459). When people are exposed to a media persona on a regular basis, they start to experience a sense of closeness, perceived friendship, and identification (Horton and Wohl, 1956).

Third, PSR could be expressed as "a one-sided relationship" between SMI and their followers. While SMI usually shares part of their lives with their followers to establish a stronger sense of sincerity and openness, most of the time, the SMIs are unaware of their followers or know nothing or just very little about the lives of all their followers (Lou and Kim, 2019; Stever, 2017). However, followers can interact with their favorite SMI; for instance, Instagram allows users to leave likes or comments, and then the influencer has the option to reply to the uploaded content. However, as influencers often receive many requests and reactions from their followers, they might be unable to genuinely respond to all of the followers' questions or comments.

One of the most popular scales focusing exclusively on measuring PSR was developed by Harmann et al. (2008), whereby PSR is conceptualized in a multidimensional structure consisting of the following main dimensions: liking, admiration, friendship (Tukachinsky and Stever, 2019), and similarity (Lou and Kim, 2019). Therefore, we adopt all these measurement dimensions as done by Lou and Kim (2019).

Since many influencers nowadays place emphasis on their credibility, the sense of parasocial relationship helps explain why influencer marketing has a persuasive appeal (Stever, 2017). Modern technologies have made social media the ideal medium for fostering parasocial relationships (Sokolova and Kefi, 2020). According to Lim and Kim (2011), parasocial relationships with influencers over social media platforms positively impact consumer attitudes and actions toward celebrity endorsement. Influencers who connect with their audience are more persuasive than unfamiliar celebrities (McCormick, 2016). A study by Hwang and Zhang (2018) revealed that parasocial relationships with SMIs enhance followers' purchase intentions.

Sincerity and TW are found to be essential characteristics of SMI to stimulate PSR with their followers (Tsai and Men, 2016). Repeated media exposure and perceived similarity and attraction are also found to be positively correlated with PSR (Bond, 2018). Another recent study suggests that parasocial relationships are developed through interactions with SMIs based on their attractiveness, similarity, and also expertise (Wang et al., 2020).

A number of research studies examined the mediating effect of parasocial relationships on the association between SMI's credibility or SI and behavioral intentions of followers as in purchase intentions (Yılmazdoğan et al., 2021; Masuda et al., 2021; Ong et al., 2021; Yuan and Lou, 2020; Lou and Kim, 2019; Reinikainen et al., 2020). Thus, we propose the following hypotheses to test the mediating effect of PSR over the association between SI and purchase intentions:

**H2**: *Source influence of social media influencers has a positive effect on the parasocial relationships between influencers and their followers.*

**H3**: *Parasocial relationship has a positive effect on the followers' intention to purchase.*

## 1.3. Brand–influencer Fit (BIF)

The concept of brand personality was first introduced in the 1960s (Saeed et al., 2022). Brand personality is defined by Aaker (1997) as "the set of human characteristics associated with the brand, "which contains aspects of brand identity (Azoulay and Kapferer, 2003). Brand personality is employed to evaluate a brand's similarity to another entity, such as in a product, another brand or an individual (Saeed et al., 2022). The tie–up between the SMI and the endorsed brand depends on the degree of perceived 'fit' between the brand image and personality and the SMI's values and standards (Tian and Li, 2022). This brand–influencer fit (BIF) represents the congruence between the SMI's endorsed message and his/her perceived image by the followers (Janssen et al., 2022; Qian and Park, 2021). Nikhashemi and Valaei (2018) propose that congruity between brand personality and consumer's personality results in positive opinions that make perceived product value outweigh its cost and, thus, yield positive intentions to purchase. Therefore, the effectiveness of the SMI's endorsement is contingent upon the degree to which the SMI's personality and expertise matches the brand personality (Von Mettenheim and Wiedmann, 2021). If media users notice a mismatch between the brand and the influencer, they cognitively recognize the incongruous association (Xie and Feng, 2022).

Product characteristics and BIF were found to affect persuasion in advertising literature (Myers, 2022; Bergkvist and Zhou, 2018). It has been found that high BIF results in higher believability compared to a lower fit (Li et al., 2022), and it positively impacts the effectiveness of advertising (Breves et al., 2019). Findings of previous research suggests that BIF has a positive effect on the followers' intention to purchase (I2P). Martínez-Lopez et al. (2020) suggest that highly commercial messages by the influencer negatively impact the trust of the followers in the influencer and also negatively impact the purchase intention. By employing the concept of brand personality to define the BIF and based on the findings in support of the positive effect of BIF on the followers' I2P, we propose the following hypothesis:

**H4**: *Brand–influencer fit has a positive effect on the followers' intention to purchase.*

SMI who actively promote products by sharing brand experiences and pictures on social media platforms provide valuable product information that passively reduces uncertainty (Berger, 1979). According to the uncertainty reduction theory (Berger and Calabrese, 1975), both parties tend to reduce their uncertainty by gathering information about the other party at the start of a relationship. Thus, the illusionary friendship with the SMIs mitigates followers' uncertainty (Lee and Lee, 2017). This explains the role of SMIs as SI in mitigating uncertainty by sharing information about the endorsed product.

The uncertainty reduction theory also posits that as relationships are developed, individuals' ability to predict the other's behavior increases. The length of time for which individuals have been acquainted for and the amount of information they acquire about others enable individuals to be more confident in their attributions about those who are known for longer periods of time (Perse and Rubin, 1989). For example, when followers develop a parasocial relationship with SMIs, over multiple episodes of media viewing, they start recognizing their area of expertise and personality traits and, thus, expect consistency and congruency with the endorsed brand personality (Schouten et al., 2019). Through parasocial relationships, the perception of BIF is formed, where BIF is assumed to be more important for SMIs than for celebrities due to the domain of expertise claimed by SMIs (Schouten et al., 2019). A study on the degree of interaction on SMIs' Instagram profile by followers and non-followers revealed a relation between a BIF and induced interactions by the SMI's followers only (Belanche et al., 2020). PSR with SMI affects brand attitudes, this is due to the fact that PSR could deepen the SMI's representation of the brand (Knoll et al., 2015).

In order to understand why people are keen to keep congruence among their cognitions, the self–consistency theory (Korman, 1974), explains that people like to keep a balance and to preserve self–image. Therefore, people tend to give favored treatment to a stimulus (product or individual) aligned with their goals and expectations (Grall and Finn, 2022). Mettenheim and Wiedmann (2021) employed opinion change and adaptation theories to explain the usefulness of product– or brand–influencer fit for adapting information processing from the source influence. An incongruence between the brand or product and the information source yields undesirable product evaluations because consumers have to change their cognitive structures (Kanungo and Pang, 1973).

In light of this discussion, it is assumed that congruence/similarity between the SMI and the followers is essential in developing and growing parasocial relationships (Mettenheim and Wiedmann, 2021; Song and Kim, 2020), which allows the transmission of information about the brand and the personality traits of the SMI over sessions of mediated viewing. Furthermore, with the followers' knowledge about the personality traits of the SMI, followers form a perception about the extent of the match between the endorsed brand and the influencer (Saeed et al., 2022; Tian and Li, 2022), and expect a congruence between the SMI and the brand (Grall and Finn, 2022; Schouten et al., 2019). Based on this discussion, we propose the following hypothesis:

***H5****: Parasocial relationship has a positive effect on the perceived brand–influencer fit.*

When the SMI's image fits the brand s/he promotes, a high level of expertise and trustworthiness is perceived by followers (Boerman et al., 2022). Some studies explain that sincerity (Lee and Eastin, 2020) and honesty (Janssen and Fransen, 2019) are important factors in consumers' judgment. Another relevant study shows that expertise and similarity with the SMI positively correlates with the brand's attitude (Dhun and Dangi, 2022). Moreover, findings from Lee and Kim (2020) showed that consumers negatively perceive promotional posting from which they think they are purely done for the money and not because the influencer is genuinely interested in the brand (Lee and Kim, 2020).

***H6****: Source influence has a positive effect on the perception of the brand–influencer fit.*

A study on the importance of examining sequential mediation dynamics of brand – related variables was conducted by Hugh et al. (2022). However, just a few studies examined BIF – related sequential mediation effects. A recent finding from Bi and Zhang (2022) affirms that brand attitude mediates the association between PSR and I2P. Furthermore, Shen (2020) confirms the sequential mediation of PSR and brand personality over the association between cuteness in marketing and the attitude toward the brand.

## 1.4.    Benign Envy (BE)

Characteristics of the audience also affect persuasion (Hovland and Weiss, 1951). Rhodes and Wood (1992) showed that people with different levels of self–esteem respond differently to the communication message; they suggest that individuals with a moderate level of self–esteem are more easily persuaded. Self–esteem is considered a suitable variable to measure individual differences of the followers (Vrontis et al., 2020). Self–esteem affects the formation of benign envy (Coopersmith, 1967). In influencer marketing, the envier (follower) compares oneself upward to the envied (SMI). Envy is viewed as a leveler with which an individual tries to level up and stabilize their emotions (Smith et al., 1999, Sayers, 1949, p.771). The follower either feels positively motivated to become like the SMI or feels negatively lacking value because of a low self–evaluation (Vogel et al., 2014).

Individual differences of self–evaluation form constructive or destructive consequences of self–esteem (Coopersmith, 1967). Differences in self-efficacy were also found to significantly and positively predict benign envy (Battle and Diab, 2022). Positive or negative emotions resulting from status comparison point to individual differences in the tendency to feel envy (Schoeck, 1969). Therefore, envy is understood as a self-relevant context (Wood, 1996). The outcome of the comparison of the envier (follower) toward the envied (SMI) can be either benign, which is associated with positive comparisons with the influencers, or malicious, where destructive evaluations materialize (Sung and Phau, 2019).

When someone wants to possess a supposedly desirable object or position that another person has, envy develops into feelings of inferiority (Parrott and Smith, 1993). Despite feeling inferior, followers who react to such comparison in such a way as to desire self-improvement and motivation experience benign envy (Van de Ven, 2012). Benign envy correlates with high hopes for success, goal setting, and better performance (Lange and Crusius, 2015). Followers develop benign envy when they feel love toward the influencers and when they want to follow their lifestyle and acquire the products they promote (Jin and Ryu, 2020).

Researchers first developed dispositional envy scales to measure an individual's malicious envy tendencies (Smith et al., 1999). Recently, researchers developed dispositional benign and malicious envy scales to differentiate the two types (Lange and Crusius, 2015). We follow the synthetization of Lange and Crusius (2015) of benign envy. They suggest that benign envy is the admiration that eventually becomes a driving factor behind people's efforts to better and enhance themselves.

Benign envy positively correlates to the intention to purchase experiential products when SMIs share an experience with followers not for showing off (Asdecker, 2022; Lin, 2018). The findings of Singh and Ang (2020) suggest that benign envy affects body–related products. In purchasing cultural capital–related products, study participants who experienced benign envy reported a greater desire to purchase (Ahn et al., 2018). The perception of benign envy toward individuals who are posting on social media increases eWOM's intention; and the effect is more substantial when the posters and reviewers have strong ties (Yan et al., 2022). The eWOM, in turn, positively affects the intention to purchase (Onofrei et al., 2022). Based on the evidence of previous research, we propose the following hypothesis:

*H7: Followers' benign envy toward social media influencers has a positive effect on the followers' intention to purchase.*

Follower–influencer relationships are often parasocial in nature (Sokolova and Kefi, 2020). Parasocial relationships are built on friendship and understanding (Chung and Cho, 2017). It is through PSR that SMIs cast an influence over their followers; for example, women who are in parasocial relationships with celebrities exhibit a desire to look like them and be part of the "club" (Greenwood et al., 2008). The illusion of friendship with the influencer is developed not only through mutual liking but also through the frequency of communication over time, which develops into an illusional form of personal ties (Tsai and Men, 2016). In the context of social media, Duan (2021) conducted a series of five studies to investigate the dynamics between tie strength, purchase intention, and benign envy. The results show that tie strength positively influences feelings of benign envy, which in turn affects the purchase intention of the displayed offering. In another study, benign and malicious envy is found to affect the intention to use (Wu and Srite, 2021). Strong ties triggers more benign envy than malicious envy (Wang et al., 2021). We thus put forward the following hypothesis:

*H8: Parasocial relationship has a positive effect on benign envy.*

Brands frequently work with SMIs since they are viewed as more credible and reputable (Bi and Zhang, 2022). Consumers' envy intensifies as SMIs' content is perceived as credible (Coelhoso et al., 2022). Followers tend to develop a more favorable purchasing attitude toward sincere endorsements of SMIs through benign envy (Lee and Eastin, 2020). In experiential products, as in travel experiences, the content of travel influencers is found to stimulate audiences' intention to travel through benign envy (Asdecker, 2022). We, thus, formulate the following hypothesis:

*H9: Source influence has a positive effect on benign envy.*

## 2. Methodology

Partial-least squares structural equation modeling (PLS–SEM) is one of the standard methods for data analysis in social science research, including marketing. It tests prediction hypotheses based on in-sample metrics that indicate how well the hypothesized model fits the data. The open–source package PLSpredict (Shmueli et al., 2016), which uses out–of–sample evaluation of the model's predictive power, has become a standard tool in the PLS–SEM analysis. However, the PLSpredict does not quantify the model's predictive performance. Toward this end, by applying ML classification algorithms, we obtain some quantifiable values of the model's predictive performance.

By comparing two independent studies, we attempt to show the importance of applying classification-based ML to quantify the model's predictive accuracy. Study – I focuses on examining the sequential mediation effect of BIF via PSR on the association between SI and the I2P, while study – II focuses on the sequential mediation effect of BE via PSR on the association between SI and I2P. The method is planned in three main stages organized in this chapter under three subsections. Figure 1 is a pictorial illustration of the method.

**Insert Figure 1 about here**

### 2.1. Stage – I: PLS–SEM Analysis

In stage – I, we evaluate the measurement and structural models using SmartPLS software. First, we start our analysis by evaluating whether the issue of common method bias is recognized. The regression-based marker technique suggests the inclusion of a marker-variable when estimating a regression equation. Common method bias is diagnosed by comparing the fit indices of two nested models, one of which connects the marker variable to the observed items of the other constructs of the study (Lindell and Whitney, 2001). This approach controls common method bias at the individual observed item-level (kock et al., 2021).

Next, we validate the measurement model, report the loadings, and follow the standard procedure to ensure convergent and discriminant validity. Regarding construct reliability, we use Cronbach's alpha coefficient (α) which measures the internal consistency, or reliability, of a set of survey items. This statistic helps determine whether a collection of items consistently measures the same characteristic. Cronbach's alpha quantifies the level of agreement on a standardized 0 to 1 scale. The acceptable threshold of $\alpha \geq 0.70$. In addition, the variance inflation factor (VIF) which detects the severity of multicollinearity by measuring the correlation among observed items is checked. The VIF values should be desirably small.

As for convergent and discriminant validity, the average variance extracted (AVE) is commonly used to assess convergent validity. AVE is a measure of the amount of variance that is captured by a construct in relation to the amount of variance due to measurement error. For interpreting convergent validity, the AVE values above 0.7 are considered very good, whereas the level of 0.5 is acceptable (Fornell and Larcker, 1981). The AVE is often used to assess discriminant validity as well. This is done based on the following "rule of thumb": the positive square root of the AVE for each of the latent variables should be higher than the highest correlation with any other latent variable. If that is the case, discriminant validity is established at the construct level. This rule is known as Fornell–Larcker criterion (Fornell and Larcker, 1981). We assess discriminant validity by (i) the square root of AVE, and (ii) the heterotrait–monotrait ratio (HTMT) of study variables which are above 0.85.

Regarding the sample's size adequacy for SEM, the Kaiser-Meyer-Olkin measure (KMO; Kaiser, 1974) is used. The KMO of sampling adequacy is the ratio of correlations and partial correlations that reflects the extent to which correlations are a function of the variance shared across all variables rather than the variance shared by particular pairs of variables. The KMO ranges from 0.00 to 1.00 and can be computed for the total correlation matrix as well as for each measured variable. Overall KMO values ≥.70 are desired (Watkins, 2018). We also check the sample size compliance with (i) the subject-to-item ratio of 10:1 (Bentler and Chou's, 1987) and (ii) the priori-sample size calculator *Daniel Soper* (Soper, 2023).

Based on theoretical foundations, we construct two structural models for each study. The first is with the sequential mediation effect of PSR with: BIF in study-I and BE in study-II. The second is constructed with the mediating effect of just PSR. Finally, we check the models' in–sample predictive power by examining the model fit indices (Hu and Bentler, 1998) and *R*–square values (Cohen, 1988). In this step, we also test the hypotheses by carefully performing mediation analyses, following the standard guidelines to evaluate mediation effects (Zhao et al., 2010; Preacher and Hayes, 2008). To assess the effect size, we evaluate the *F*–Square values; *F*-Square is the change in *R*-Square when an exogenous variable is removed from the model, whereas *R*-square represents the proportion of the variance for a dependent variable that is explained by an independent variable or variables in a regression model. The *F*-Square values of all variables are evaluated at the significance level ($\rho$-value ≤ 0.05). If the *F*-Square value is ≥ 0.35, the effect size is considered large, if ≥ 0.20, the effect size is medium, if ≥ 0.15, the effect size is small. An *F*-Square value ≤ 0.15 is considered trivial and the hypothesis is rejected despite the significant $\rho$-value (Cohen, 1988).

To validate the advantage of adopting a structural model instead of a simple linear model (LM), we use the PLSpredict tool to assess the predictive power from an out–of–sample perspective (Shmueli et al., 2016). The PLSpredict algorithm employs k–folds cross–validated cases to report root mean square error (RMSE), mean absolute error (MAE), and mean absolute percentage error (MAPE) prediction errors. We set the cross–validation to ten folds (k=10) with the number of repetitions (r =30) to obtain the following two naïve benchmarks from the PLSpredict: 1) linear model predictions and 2) mean value *Q*–Square to measure the predictive quality of the path estimations. Regarding the predictive relevance of the latent variables, the *Q*–Square values should be above zero for the model to prove predictive relevance. The Calculation for *R*-Square and *Q*-Square are almost identical, the *R*-Square is computed as 1− [residual sum of squares (RSS)/total sum of squares (TSS)], while the *Q*-Square is computed as 1− [predicted residual error sum of squares (PRESS)/ TSS]. The only difference being that RSS is calculated from the data on which the algorithm is trained, and PRESS is calculated from the held-out data.

Following the guidelines for reporting the results for PLSpredict presented by Shmueli et al. (2019), we compare the structural model of the PLS–SEM and the LM to examine their predictive errors. Sarstedt et al. (2021) recommends using MAE to assess the prediction errors when the distribution of the prediction errors is highly asymmetric; otherwise, RMSE is recommended. We also compare the *Q*–Square values of the observed items of the endogenous constructs in the PLS model vs. the simple LM (Shmueli et al., 2019).

We use the bootstrap procedure by considering 5,000 subsamples with replacement and 20 iterations to determine the significance of the mediation relationships (Carrión et al., 2017). We use the *t*–value and its corresponding $\rho$-value as indicators to accept or reject the hypotheses. For each test, the *t*-value is a way to quantify the difference between the population means and the $\rho$-value is the probability of obtaining a *t*-value with an absolute value at least as large as the one we observed in the sample data if the null hypothesis is actually true. We follow the guidelines of Demming et al. (2017); Hayes (2013); and Preacher and Hayes (2008) for reporting the results of the mediation analysis.

## 2.2.    Stage – II: Obtaining Predictive Accuracy through ML Classification

The numerical dependent variable is transformed into a categorical variable in classification analysis. Therefore, we transform the numerical dependent variable into binary. The threshold for classifying the dependent variable into binary is found by slicing the dependent variable into deciles and assigning the value 1 when the numerical value is greater than or equal to the decile threshold value and 0, otherwise. The decile threshold value corresponding to the best–performing classifier is used for classification. We intend to evaluate the hypothesis of whether the independent variable predicts the dependent variable with an acceptable accuracy level. For this purpose, six classification–based ML algorithms are used with 10–fold cross–validation technique to assess the prediction accuracy of the structural model. We use the most common classification algorithms that belong to the following classification categories:

(1) Bayesian Network under the Bayesian classification approach
(2) Logistic regression under the functions' classifiers
(3) Local weighted learning (LWL) under the lazy classification approach
(4) Adaptive boosting M1 under the meta classification
(5) One Rule under rules classification category
(6) J48 classification algorithm under the decision trees

In ML classification, the confusion matrix assesses the model's performance (Japkowicz and Shah, 2011). Several metrics are computed from the confusion matrix, such as accuracy, precision, and recall. Precision quantifies the number of positive class predictions that actually belong to the positive class. In contrast, recall or sensitivity quantifies the number of positive class predictions made out of all positive examples in the dataset (Powers, 2011). Precision is thus a measure of quality, while recall is a measure of quantity. The F–measure provides a single score that balances both the concerns of precision and recall in a single score. Accuracy gives the model's overall accuracy, meaning the proportion of the total data points correctly classified by the classifier.

The best–performing algorithm is selected by evaluating the accuracy, precision, recall, true and false positives rates resulting from the confusion matrix of the classifier. We follow the rule of thumb of identifying the best-performing model through its maximum precision subject to the recall of 0.75 (e.g., Sharkasi et al., 2015). A general rule for evaluating accuracy scores is that over 90% is considered very good, 70% to 90% is considered good, and 60% to 70% is just okay (Stehman, 1997).

## 2.3. Stage – III: Comparative Prediction Performance

We compare prediction performance to assess the role of BIF and BE in predicting I2P via PSR. We plan comparative analyses in three main steps:

Step – I: comparing the mediation effects

Step – II: evaluation of predictive power and accuracy

Step – III: validation of results; we apply the split–test technique to compare possible sample bias.

### 2.3.1. Step – I: Comparing Mediation Effects

We use PLS–SEM to investigate the sequential mediation of (i) BIF with PSR, and (ii) BE with PSR on the association between SI and I2P, denoted by SI→I2P, in study – I and – II, respectively. We carefully follow the standard guidelines for evaluating mediation effects (Demming et al., 2017; Hayes, 2013; Zhao et al., 2010; Preacher and Hayes, 2008). We use the bootstrap procedure by considering 5,000 subsamples with replacement and 20 repetitions to determine the significance of the mediating relationships (Carrión et al., 2017). The sequential mediation model consists of the following main paths:

(i)   The effect of SI on the I2P through PSR in both studies, denoted by SI → PSR → I2P.

(ii)  The effect of PSR on I2P through: BIF in study – I, and BE in study – II, denoted by PSR→BIF→I2P and PSR→BE→I2P, respectively.

(iii) the mediation effect of PSR on: SI→BIF, in study – I, and SI→BE in study – II, denoted by SI→PSR→BIF and SI→PSR→BE, respectively.

(iv)  the sequential mediation of PSR with: BIF on SI→I2P in study – I, and with BE on SI→I2P in study– II, denoted by SI→PSR→BIF→I2P and SI→PSR→BE→I2P, respectively.

We conduct within–study comparative analysis by assessing the sequential mediation effect of PSR with BIF on SI→I2P vs. the mediation effect of just PSR on SI→I2P. Similarly, in study – II, we compare the sequential mediation effect of PSR with BE on SI→I2P vs. the mediation effect of just PSR on SI→I2P.

For comparing the mediation effects, we rely on two main metrics, first, the absolute difference between the total effects (TE) and the direct effects (DE). Second, the ratio of the indirect effect (IE) over the TE (Agler and De Boeck, 2017). With the intervening effect of the mediators, the total effect is divided into DE and IE. Thus, it is expected for the DE to be lower than the TE in the presence of a significant mediator, explaining IE. The wider the gap between TE and DE, the stronger the role of the mediator in the relationship. The difference (TE – DE) quantifies the indirect effect's contribution to the prediction of the dependent variable, I2P. We also compare the mediation effects between–studies to assess which sequential mediation effect has a higher role in predicting I2P.

### 2.3.2. Step – II: Evaluation of Predictive Power and Accuracy

In step – II of stage – III, we conduct within–study and between–studies comparisons. For this purpose, we evaluate:

I.  Predictive accuracy resulting from the confusion matrix of the best–performing classification algorithm.

II. Predictive and descriptive power resulting from the measurement and structural models of the PLS–SEM analysis. For this, we consider the following:

1.  The total variance explained resulting from the measurement model. The higher the percentage of the total variance explained, the higher the percentage of the provided information. Hence, this measure explains how much variation in the dataset is attributed to the model's total variance.

2.  The model selection criteria, Akaike Information Criterion (AIC) obtained from the measurement model. This criterion is commonly used to compare different models and determine the best fit for the data. It is a way to measure the goodness–of–fit and penalizes the model for over–fitting.

3.  Assessing the descriptive power of the model by evaluating its goodness–of–fit (Sarstedt and Danks, 2022; Shmueli, 2010, p.3). Various in–sample metrics are used to indicate how well the hypothesized model fits the data. We evaluate the global fit indicator as the standardized root mean square residual (SRMR), which measures the model's approximate fit (Hu and Bentler, 1998). We follow Brown's (2015) recommendation that the model fit in SEM is also assessed by evaluating the normalized fit index (NFI). Therefore, we consider the model fit indicators as SRMR, and NFI for comparisons of descriptive power (Sarstedt and Danks, 2022).

4.  The change rate in average RMSE values of the observed items of I2P. The RMSE errors obtained from *PLSpredict* are two types, one for the PLS and another for the linear model, denoted by $RMSE_{PLS}$ and $RMSE_{LM}$, respectively. The PLS and LM evaluate the predictions with and without the structural model, respectively (Shmueli et al., 2016). To verify the structural model's predictive power, the $RMSE_{PLS}$ values for all or most observed items should be lower than their corresponding $RMSE_{LM}$, which indicates the advantage of adopting the structural model compared to the simple LM. The PLSpredict evaluates the RMSE value for each observed item of the endogenous variables. For comparative purposes, we evaluate the average of $RMSE_{PLS}$ and $RMSE_{LM}$ scores of the observed items of the dependent variable, I2P, and then compute the difference between the averages, ($RMSE_{LM} - RMSE_{PLS}$).

To standardize these differences, we consider the percentage difference for within-study comparisons of the sequential mediation compared to the mediation of just PSR. The percentage difference (Cole and Altman, 2017) is computed by taking the absolute difference between the two scores divided by their average as

$$Percent\ difference\ for\ the\ i-th\ observed\ item = \left|\frac{RMSE_{LM-i} - RMSE_{PLS-i}}{\left(\frac{RMSE_{LM-i} + RMSE_{PLS-i}}{2}\right)}\right| \times 100\%\ .$$

The percentage difference is widely used in economics, for example, in computing the elasticity of demand on a non–linear curve (Elbasha, 1997). In addition, it is used when we compare values that relate to one another or have a similar nature in uncertain situations (Alvis and Abdelkefi, 2023).

For between–studies comparisons of the criteria listed in $(1-4)$, we use the change rate computed by taking the difference between the two values divided by the latest. For example, to compute the change rate of the total variance explained for the sequential mediation model, we compute the change rate as follows:

$$change\ rate = \frac{|\ Variance_{study-I} - Variance_{study-II}\ |}{Variance_{study-II}} \times 100\%\ .$$

### 2.3.3.   Step – III: Validation of results

The third and last step of stage – III is validating results. To eliminate the possibility that the sample of either study has a role in steering results in favor of a particular variable, we use the split–test technique for comparing the performance of the sample in predicting I2P. We utilize the PLS models with only the common mediator, PSR, so that we could use the sample of study – I as a training dataset, and study – II as a testing dataset and vice versa and then assess the resulting predictive errors.

## 3.   Study – I: The Effect of Brand–influencer Fit (BIF)

### 3.1.     Measures, Instrument Design and Sampling of Study – I

The independent second-order latent variable, SI, is defined by the first-order latent variables: EXP, TW, ATT, and SIM which are measured by the corresponding values of the observed items: EXP–1⋯EXP–4, TW–1⋯TW–4, ATT–1⋯ATT–4, and SIM–1⋯SIM–3, respectively, as shown in Table 1. The independent latent variable, PSR, is measured by the corresponding values of the observed items: PSR–1⋯PSR–5 (see Table 1). In addition, this model incorporates the independent variable, BIF, measured by the corresponding values of the observed items: BIF–1⋯BIF–5 (see Table 1). Finally, the dependent latent variable, I2P, is measured by the corresponding values of the observed items: I2P–1⋯I2P–5 (see Table 1).

---

**Insert Table 1 about here**

---

The questionnaire consists of four main sections, section one introduces the study, explains the objectives, and clearly describes the researchers' ethical obligation towards privacy and data handling. Participants had to sign a consent form to confirm their voluntary participation in the study before they could access the questionnaire. In addition, this section verifies the participants' eligibility to participate in the study. Participants should be active on one of the most popular social media platforms, YouTube, Facebook, TikTok, Instagram, Twitter, and Pinterest. Statcounter (2022) states that the top five most popular social media platforms worldwide are Facebook, Twitter, Instagram, Pinterest, and YouTube. Thus, we consider these platforms for our study. We also enabled the participants to list other social media platforms they use. Active participants are identified as (i) being active on social media for more than three hours weekly and (ii) following at least one influencer on Instagram.

Demographic data are collected in the second section of the questionnaire; participants present their gender, age, nationality or country of origin. They also have to name the social media platform they are active on the most, and the average weekly hours spent on it. The third section of the questionnaire includes a list of questions to capture the observed items of the study described in Table 1. A seven-point anchored scale is used to capture the participants' responses. The anchored scale represents one as strongly disagree and seven as strongly agree. We adopt the questions and scale from the list of references in Table 1. We follow Lou and Kim (2019) for synthesizing SI and measuring I2P and Breves et al. (2019) for defining the formation of perceived BIF. The last part of the questionnaire includes three questions related to the participants' online shopping habits but unrelated to the study to be used as a marker variable to test for *Common Method Variance*.

The data were collected in two stages online following the *Convenience Sampling Method*. The researchers posted the questionnaire through social media and emails to potential participants in Europe and Southeast Asia (SEA). During Jun – Aug 2022, a total of 280 completed questionnaires were collected from European participants. During the subsequent months of Nov–Dec 2022, a total of 131 questionnaires were collected from participants in SEA using the same questionnaire and protocol. All completed questionnaires are clear of biased responses. The description of the sample profile is reported in Table 2.

---

**Insert Table 2 about here**

---

The majority of the samples are females (60.58%). About 80% of the samples are young individuals aged 18 –27. Regarding the average hours spent on social media platforms, about 83% of the samples were active on Instagram, followed by YouTube (68.61%), Facebook (57.66%), TikTok (26.52%), Twitter (23.36%), and Pinterest (16.06%).

## 3.2.    Measurement Model of Study – I

First, we construct a marker-variable for a common method bias diagnosis. The marketer-variable is conceptually unrelated to other factors in the study. Hence, in this study, it is measured by the common characteristics of online shoppers, such as self-expression, selectivity, and deal driven (Parment, 2013). Comparing the fit indices of the nested models, our model does not raise concerns of common method bias. Moreover, the variance of any single construct was up to 13.322% of the total variance explained, which amounts to 47.578%. Regarding the sample's size adequacy, we evaluated the KMO ratio for sampling adequacy as 0.944 with a significant chi–square (<0.001). A total number of 411 questionnaires is considered an adequate sample size for this study, since it complies with the subject-to-item ratio of 10:1 (Bentler and Chou's, 1987) and with the priori-sample size calculator *Daniel Soper* which suggests the estimated minimum sample size not to be less than 250.

Table 3 shows the standardized loadings of the measurement model's observed items or indicators on their respective latent variables. All standardized loadings are above 0.75. Regarding construct reliability, we report the Cronbach's alpha (α) scores of 0.925, 0.957, 0.863, 0.863, 0.879, 0.900, 0.895 for the variables: EXP, TW, ATT, SIM, SI, PSR, BIF, and I2P, respectively, which are above the 0.70 threshold. In addition, the composite reliability (CR) scores of the previously listed variables are above the 0.70 threshold: 0.947, 0.969, 0.906, 0.916, 0.909, 0.926, and 0.934, respectively. These results indicate well–constructed observed items that reflect their correlation to their respective latent variables. Moreover, all VIF scores are less than 2.90, thus satisfying the 3.0 threshold (Hair et al., 2021). This is an indication of no multicollinearity issues.

**Insert Table 3 about here**

As for convergent and discriminant validity, as shown in Table 3, all AVE values are higher than the 0.5 threshold. Moreover, as shown in Table 4, the square root of any AVE value is higher than the highest correlation with any other latent variable. In addition, the HTMT values are above the 0.85 threshold. The findings support the convergent and discriminant validity of the study's constructs.

**Insert Table 4 about here**

### 3.3. Structural Model Not Including Brand–influencer Fit (BIF)

The evaluation of the outer measurement model indicates good reliability and validity. We could now assess the inner structural model by considering the role of the mediator variable, PSR, on the association of SI→I2P. Not including BIF, the KMO of our model is 0.946, and the total variance explained is 44.374%. Figure 2 shows the analysis output of *SmartPLS* 3.0 software. The path coefficients between SI and EXP, TW, ATT, SIM are found at the significance level <0.05 as: 0.320 with 30.24 *t*-value, 0.360 with 33.98 *t*-value, 0.253 with 30.23 *t*-value, 0.232 with 34.68 *t*-value, respectively. The path coefficients between the latent variables of hypotheses H1–H3 and their corresponding $p$-values are shown in Figure 2.

---

**Insert Figure 2 about here**

---

Following Hu and Bentler (1998), we evaluate the standardized-root-mean-square-residual (SRMR) as 0.073; this value is below the maximum acceptable 0.08 threshold. We also evaluate the normalized fit index (NFI) value of our model as 0.920, which is acceptable (Bentler and Bonett, 1980). Thus, no evidence of model misspecification is detected (Bentler and Bonett, 1980).

Table 5 reports the estimated standardized coefficients and their significance for testing hypotheses. All the hypotheses are accepted with a significant $p$-value $\leq 0.05$. To assess the effect size, we evaluate the *F*–Square values. The path coefficients are evaluated as:

- $\beta = 0.227$ with *F*–Square= 0.229 as medium effect for SI → I2P
- $\beta = 0.786$ with *F*–Square= 0.354 as large effect for SI → PSR
- $\beta = 0.238$ with *F*–Square = 0.302 as medium effect for PSR → I2P

The *R*–Square values are all above 0.26 (see Table 5) which suggests strong statistical power (Cohen, 1988). Regarding the predictive relevance of the latent variables, the *Q*–Square values are all evaluated above zero (see Table 5); accordingly, the model has a predictive relevance.

---

**Insert Table 5 about here**

---

To assess the mediation role of PSR on the association SI→I2P, we evaluate the following:
- The path coefficient of the indirect effect over SI→I2P is found as $\beta = 0.177$ with *t*–value = 17.354 and the corresponding estimated significance level $p$–value= 0.001.
- The coefficient of the total effect of SI→I2P is found as $\beta = 0.400$ with *t*–value =18.694 and the corresponding $p$–value= 0.00.

- The path coefficient of the direct effect of SI→ I2P was found as $\beta$= 0.226 with $t$–value =17.304 and the corresponding $p$–value= 0.001. This coefficient is less than the corresponding coefficient of the total effect of SI→I2P, this is due to the mediation role of PSR.

This indicates that PSR partially mediates the relationship between SI→ I2P, and that the mediation is positive complementary because all coefficients are positive (Demming et al., 2017). Regarding the predictive power analysis using *PLSpredict*, since the distribution of the error terms are not highly asymmetric, we evaluate RMSE to confirm the privilege of the structural model (Sarstedt et al., 2021). The decreased rate of the PLS's errors to the LM for the observed items of the latent variable: I2P–1, I2P–2, I2P–3, are: 1.45%, 0.00%, and 1.00%, respectively (see Table 6). Similarly, the computed decreased percentage of RMSE for the observed items of PSR were all positive, indicating higher error terms of the LM compared to the PLS model. This is an indication of an acceptable predictive power of the PLS model.

In addition, the $Q$–square values of the PLS model are higher than the LM except for one item, I2P–2, where the values were equal. Despite the equal RMSE and $Q$–square values in the PLS and LM for the observed item I2P–2, the overall results indicate an acceptable predictive power of the PLS model. With these observations, we confirm the privilege of the outlined structure of the PLS model with the mediating effect of PSR.

---

**Insert Table 6 about here**

---

## 3.4. Prediction Accuracy of the model not Including Brand–influencer Fit

We aim to quantify the model's predictive performance; to achieve this, the latent response variable is transformed into a binary form with six classification–based ML algorithms to obtain the model's accuracy score. Table 7 reports the performance metrics of the confusion matrix of the highest–performing classifier for each path in the structural model.

---

**Insert Table 7 about here**

---

The predictive accuracy scores for the mediation of PSR on the association SI→I2P through: *H1*: SI→I2P; *H2*: IS→PSR; and *H3*: PSR→I2P, are evaluated as: 64.23%, 77.86%, and 64.48%, respectively. All classifier algorithms for all hypotheses have produced the same accuracy, recall and precision scores except for hypothesis *H1* where AdaBoostM1 and J48 algorithms provide lower accuracy scores. The results support the importance of the mediation of PSR since the prediction accuracy, recall and precision were all over 60%. The prediction accuracy of the model with all associations is 64.48%, with recall and precision values over 60%, and TP rate of 0.645, which is much higher than FP rate of 0.384. This prediction performance is considered just okay.

### 3.5. Sequential Mediation: Structural Model Including Brand–influencer Fit (BIF)

In this part, we study the extension of the model in subsection 4.3. by considering the role of the mediator variables, PSR and BIF on the association of SI→I2P. The latent variables: EXP, TW, ATT, and SIM, loaded significantly on the second–order variable, SI, with the corresponding coefficient values: 0.320 with 36.69 $t$-value, 0.360 with 33.53 $t$-value, 0.253 with 31.16 $t$-value, and 0.232 with 33.60 $t$-value, respectively (see Figure 3). To test hypotheses *H1*, *H2–H3*, and *H4–H6*, presented in subsections 2.1, 2.2, and 2.3, respectively. The path coefficients and their corresponding $\rho$–values are shown in Figure 3.

---

**Insert Figure 3 about here**

---

Table 5 displays the estimated standardized coefficients and their significance levels. All the hypotheses are accepted with a significant $\rho$–value $\leq 0.05$. All $F$–Square values are evaluated at the significance level $\rho$–value $\leq 0.05$. The $F$–Square and $R$–Square values are all evaluated above 0.20 and 0.25, respectively, hence, they are considered acceptable (Cohen, 1988). The path coefficients of the accepted hypotheses, *H1 – H5,* of the model are evaluated as:

- $\beta$= 0.150 with $F$–Square= 0.290 as a medium effect for SI→I2P
- $\beta$= 0.740 with $F$–Square= 0.421 as a large for SI→PSR
- $\beta$= 0.234 with $F$–Square=0.231 as a medium effect for PSR→I2P
- $\beta$= 0.522 with $F$–Square=0.374 as a large effect for PSR→BIF
- $\beta$= 0.119 with $F$–Square=0. 0.290 as a medium effect for BIF→I2P

Regarding the rejected hypothesis, *H6*, the path coefficient of SI on BIF is evaluated as $\beta$=0.120, but the relationship was found insignificant with $t$–value =1.678 with $\rho$–value=0.510. The $R$-Square values are all above 0.26 (see Table 5), which suggests strong statistical power (Cohen, 1988).

As per the global fit indices, following Hu and Bentler (1998), the SRMR value is evaluated as 0.072, which is below the acceptable threshold of 0.080. The NFI value is evaluated as 0.930, which is below the acceptable threshold of 0.90. Thus, no evidence of model misspecification is detected (Bentler and Bonett, 1980).

To assess the mediation role of BIF via PSR on the association SI→I2P, we first evaluate the mediation role of PSR and BIF on SI→I2P:

- The path coefficient of the indirect effect of SI→I2P through the mediator variable, PSR, is found as $\beta$=0.173 with $t$–value = 17.354 and the corresponding estimated $\rho$–value= 0.001.
- The path coefficient of the indirect effect of the sequential mediation of BIF via PSR, denoted by SI→ PSR → BIF →I2P, is found as $\beta$= 0.046 with $t$–value =17.978 and the corresponding $\rho$–value= 0.042.
- The coefficient of the total effect of SI→I2P is found as $\beta$= 0.369 with $t$–value =17.124 and the corresponding $\rho$–value= 0.000.
- The path coefficient of the direct effect of SI→ I2P is found as $\beta$= 0.150 with $t$–value =17.717 and the corresponding estimated $\rho$–value= 0.047, which is weaker than the total effect of SI→I2P, this is due to the sequential mediating role of PSR with BIF.

These mediation effects indicate that the relationship between SI and I2P is fully mediated by PSR and BIF, and that the mediation is positive complementary. Second, we evaluate the mediation role of PSR on SI→BIF by:

- The path coefficient of the indirect effect of the mediator variable, PSR, on SI→BIF is found as $\beta$= 0.286 with $t$–value=18.424 and the corresponding $\rho$–value= 0.000.
- The coefficient of the total effect of SI→BIF is found as $\beta$= 0.386 with $t$–value =18.42 and the corresponding estimated $\rho$–value= 0.000.
- The path coefficient of the direct effect of SI→ BIF is found as $\beta$= 0.120 with $t$–value =1.678 and the corresponding estimated $\rho$–value= 0.510, which causes to reject *H6* hypothesis.

Since hypothesis *H6* is rejected, the relationship between SI and BIF is partially mediated by PSR, and that the mediation is positive complementary. Third, we evaluate the mediation role of BIF on PSR→I2P:

- The path coefficient of the indirect effect of the mediator variable, BIF, on PSR→I2P is found as $\beta$= 0.062 with $t$–value = 17.718 and the corresponding estimated $\rho$–value= 0.043.
- The coefficient of the total effect of PSR→I2P is found as $\beta$= 0.296 with $t$–value =18.545 and the corresponding estimated $\rho$–value= 0.000.
- The path coefficient of the direct effect of PSR→I2P is found as $\beta$= 0.234 with $t$–value =18.02 and the corresponding estimated $\rho$–value= 0.000, which is less than the total effect of PSR→I2P, this is due to the mediating role of BIF on this association.

This indicates that the relationship between PSR and I2P is fully mediated by BIF, and that the mediation is positive complementary. In summary, the sum of all coefficients of the indirect effects of this model is 0.667 (0.173+0.046+0.386+0.062). While the sum of all direct effects on I2P is evaluated as 0.384 (0.15+0.234). The sum of direct and indirect effects is equivalent to the sum of all total effects evaluated as 1.051 (0.369+0.386+0.296).

---

**Insert Table 8 about here**

---

The results reported in Table 6 indicate the importance of the PLS structural model in reducing prediction errors compared to the LM. The RMSE reductions of the PLS model compared to the LM for the observed items of the dependent variable, I2P–1, I2P2, I2P– 3, are at the rates: 1.49%, 1.14%, and 0.13%, respectively (see Table 6). Also, the RMSE reductions of the PLS model compared to the LM for the observed items of the independent variables, PSR and BIF, are all positive, indicating higher error terms of the LM compared to the PLS model (see Table 6). Thus, the predictive power of the PLS model is acceptable. Regarding predictive relevance, the $Q$–Square values are higher for the PLS model compared to the LM, which indicates the privilege of the PLS model.

### 3.6. Sequential Mediation: Prediction Accuracy Including Brand–influencer Fit (BIF)

We follow the same steps applied in section 5.4. and use the same classification algorithms with 10–fold cross–validation technique to quantify the predictive performance of the sequential mediation model. Table 7 reports the performance metrics of the resulting confusion matrix. The prediction accuracy of $H1$: SI→I2P, $H2$: IS→PSR, and $H3$: PSR→I2P are 64.23%, 77.86%, and 64.48%, respectively. For $H4$: PSR→BIF, and $H5$: BIF→I2P, the prediction accuracy scores are 66.42%, and 63.02%, respectively. These scores show an acceptable prediction accuracy scores for $H1$, $H3$, $H4$ and $H5$, and a good accuracy prediction for $H2$.

All classifiers provide the same accuracy, precision, and recall scores for all hypotheses except for hypothesis $H1$, the AdaBoost M1 and J48 provide lower accuracy scores. Since all accuracy scores are over 60%, the results support the importance of the mediation effect of: (i) PSR through SI→PSR, (ii) the mediation of BIF through PSR →BIF, and the (iii) sequential mediation of BIF via PSR on the association of SI→I2P. With the inclusion of BIF, the model's prediction accuracy with all associations is evaluated as 70.18%, which is enhanced compared to the 64.48% accuracy score of the model not including BIF. Therefore, it is recommended to consider the sequential mediation of BIF via PSR on the association of SI→I2P instead of just the mediation effect of PSR.

## 4. Study – II: The Effect of Benign Envy – Analysis and Results

## 4.1.     Measures, Instrument Design and Sampling of Study – II

In this study, we also examine the mediation effect of just PSR on the association of SI→I2P and the sequential mediation of BE via PSR on SI→I2P. The variables, SI, PSR, and I2P are defined as in study – I (see Table 1). Study – II incorporates the independent variable, BE instead of BIF, which is defined by eight observed items, BE–1 ⋯ BE–8, described in Table 1. Lange and Crusius (2015) considered the same observed items for measuring BE.

The questionnaire of Study – II consists of four main sections. Section one introduces the study by explaining the objectives and clearly describing the ethical obligation of researchers towards privacy and data handling. Furthermore, it verifies the participants' eligibility to participate in the study. Participants should be (i) active on Instagram for at least 3 hours a week, (ii) have at least one favorite influencer they follow on Instagram, and (iii) view the influencer's content almost every time they are on the platform.

Participants must sign a consent form to confirm their voluntary participation in the study. Upon submitting the consent form, participants continue to the second section of the questionnaire, which consists of demographic questions where participants provide their age, gender and their average weekly time spent on Instagram. The third section of the questionnaire includes a list of questions to capture the observed items of this study. A seven–point anchored scale is used to capture the participants' responses. The anchored scale represents one as 'strongly disagree' through seven as favorite 'strongly agree'. Participants are asked to answer questions in relation to their Instagram influencer. The fourth section asks some questions to construct a marker-variable for evaluating common method bias, listed in subsection 5.2.

The data was collected using the convenience sampling method by sending social media and email messages to students at an international University in Tokyo city. The data was collected during Jun–Aug 2022. A total of 360 completed questionnaires were collected. we checked all completed questionnaires for biased responses and five questionnaires were eliminated. A 355 data points are considered, and the sample size is adequate since it complies with the subject–to–item ratio of 10:1 (Bentler and Chou, 1987) and the priori–sample size calculator, *Daniel Soper*, suggesting that the sample size not to be less than 300. The description of the sample profile is reported in Table 9.

---

**Insert Table 9 about here**

---

The majority of the samples are females (67.61%). Over 90% of the samples consist of young individuals aged 17–26. Regarding the average hours spent on Instagram, about 62% of the samples spend from 3 to 5 hours weekly on this social media platform, and about 14% spend more than seven hours weekly on it.

## 4.2. Measurement Model of Study – II

To test common method bias, we apply the Ad–hoc marker–variable technique (Lindell and Whitney, 2001). We construct a marker-variable by the following questions (i) I feel bad when an influencer has a lot more than me, (ii) I feel disadvantaged when I cannot buy products sponsored by an influencer, (iii) I feel hatred towards an influencer when her/his content pulls me down. As explained in section 4.2, we regress the CLF to the marker-variable and evaluate the path coefficients of the CLF model in regard to the added marker-variable. The evaluated coefficients are below the basic CLF. Moreover, each factor in the model contributes up to 15.039% of the variance, where their total contribution to the variance amounts to 48.51%. Thus, it is more likely to have a regression model with less model method bias. Regarding the sample's size adequacy for SEM, we evaluate the Kaiser–Meyer–Olkin ratio as 0.969 with a significant chi-square (<0.001).

Table 10 shows the standardized loadings of the observed items of the measurement model on their respective latent variables. All standardized loadings are above 0.70. Regarding construct reliability, the Cronbach's alpha (α) coefficients are evaluated as 0.846, 0.863, 0.731, 0.834, 0.840, 0.886, and 0.923 for the variables: EXP, TW, ATT, SIM, SI, PSR, BE, and I2P, respectively, which are all above the 0.70 threshold (see Table 10). In addition, the CR scores of the previously listed variables are evaluated as 0.896, 0.907, 0.832, 0.900, 0.886, 0.929, and 0.937, respectively, which are all above the 0.70 threshold as well (see Table 10). These results indicate that the observed items are well constructed. In addition, the VIF values for all observed items are less than 2.60, which satisfies the 3.0 threshold necessary to exclude multicollinearity.

---

**Insert Table 10 about here**

---

Regarding convergent validity, the AVE values are all higher than the 0.50 threshold (Fornell and Larcker, 1981). As per discriminant validity, the square root of AVE of each construct is larger than the corresponding correlations with each of the remaining constructs. We also assess whether the HTMT of the variables are less than the 0.85 threshold. The findings reported in Table 11, support the convergent and discriminant validity of the constructs.

---

**Insert Table 11 about here**

---

## 4.3.    The Structural model not Including Benign Envy (BE)

In this section, we assess the inner structural model by considering the role of the mediator variable, PSR, on the association of SI→I2P without considering BE. The KMO of this model is evaluated as 0.961, and the total variance explained is 42.374%. Figure 4 presents the corresponding inner structural model. The path coefficients between SI and EXP, TW, ATT, and SIM are evaluated at the significance level (<0.05) as 0.330 with 33.12 $t$-value, 0.361 with 28.17 $t$-value, 0.195 with 14.36 $t$-value, and 0.269 with 26.60 $t$-value, respectively. The path coefficients between the latent variables of the hypotheses *H1 – H3* and their corresponding $\rho$-values are shown in Figure 4.

---
### Insert Figure 4 about here
---

Following Hu and Bentler (1998), we evaluate the SRMR as 0.075, this value is below the maximum acceptable 0.080 threshold. We also evaluate the NFI as 0.920, which is greater than the acceptable 0.90 threshold. Thus, the structural model fit indices are all acceptable, and no evidence of model misspecification is detected.

Table 12 shows the estimated standardized coefficients and their significance for testing the hypotheses. All the hypotheses, *H1 – H3,* are accepted with a significant $\rho$-value $\leq 0.05$. To assess the effect size, we evaluate the $F$–Square values. The $F$–Square values of all constructs are evaluated at the significance $\rho$–value $\leq 0.05$ and are all above 0.20. The path coefficients are evaluated as:

- $\beta$= 0.518 with $F$–Square= 0.549 as large effect for SI→I2P

- $\beta$= 0.815 with $F$–Square =0.857 as large effect for SI→PSR

- $\beta$= 0.323 with $F$–Square = 0.103 as small effect for PSR→I2P

The $R$–Square values are all above 0.26 (see Table 12) which suggests a strong statistical power (Cohen, 1988). Regarding the predictive relevance of the latent variables, the $Q$–Square values are all evaluated above zero (see Table 12). So, the model has a predictive relevance.

---
### Insert Table 12 about here
---

To assess the mediation role of PSR on the association SI→I2P, we evaluate:

- The path coefficient of the indirect effect of SI→I2P through the mediator variable PSR, which is found as $\beta$=0.268 with $t$–value = 19.022 and $\rho$–value= 0.000

- The path coefficient of the total effect of SI→I2P is found as $\beta$=0.779 with $t$–value=28.761 and $\rho$–value=0.000

- The path coefficient of the direct effect of SI→I2P is found as $\beta$=0.512 with $t$–value=17.40 and $\rho$–value= 0.000, which is less than the corresponding coefficient of the total effect of SI → I2P, this is due to the mediating role of PSR.

This indicates that PSR partially mediates the relationship between SI and I2P and that the mediation is positive complementary because all coefficients are positive (Demming et al., 2017).

To confirm the privilege of the structural model, we use the open–source package, *PLSpredict* of *SmartPLS* 3.0 software. Following Sarstedt et al. (2021), we use RMSE and not the MAE to assess the prediction errors since the distribution of the error terms are not highly asymmetric. Following Shmueli et al. (2019) for reporting the results of *PLSpredict*, we compare the structural model of the PLS–SEM and the LM to examine their predictive errors. We also compare the $Q$-Square values of the observed items of the endogenous latent variable in the PLS model vs. the simple LM. The *PLSpredict* analysis indicates the importance of the structural model in reducing the prediction errors compared to LM. The decreased rate of the PLS errors to the LM errors for the observed items of the latent variables I2P–1, I2P–2, I2P–3, are: 0.57%, 0.10%, and 0.48%, respectively (see Table 13). Similarly, the computed decreased percentage RMSE for the observed items of PSR are all positive, indicating higher error terms of the LM compared to the PLS model. In addition, all the $Q$–Square values of the PLS model are higher than the LM values. So, we conclude the privilege of the outlined structure of the PLS model with the mediation effect of PSR compared to the LM, and its predictive power.

**Insert Table 13 about here**

## 4.4. Prediction Accuracy of the model not Including Benign Envy (BE)

We quantify the model's predictive performance by transforming the latent response variable into binary with six classification–based ML algorithms to compare the accuracy score. Table 14 reports the performance metrics of the confusion matrix of the highest–performing classifiers for each path in the structural model.

**Insert Table 14 about here**

The predictive accuracy of the mediation of just PSR on the association SI→I2P through: *H1*: SI→I2P; *H2*: IS→PSR; and *H3*: PSR→I2P are evaluated as 68.17%, 72.11%, and 72.68%, respectively (see Table 14). Among the six different classification–based ML algorithms, the AdaBoost M1 algorithm produced the highest accuracy for *H1*. For *H2* and *H3*, all algorithms provided the same accuracy, recall, and precision scores, except for AdaBoost M1, which performed slightly lower. The results support the acceptance of the mediation role of PSR through SI→I2P as the accuracy of hypotheses *H1 – H3* are all over 60% (see Table 14). The prediction accuracy of the model with all associations is at least 68.17%, with recall and precision values over 60%, and TP rate of 0.682, which is much higher than FP rate of 0.360. This prediction performance is considered acceptable.

## 4.5. Sequential Mediation: Structural model Including Benign Envy (BE)

In this section, we study the extension of the model in subsection 5.3 by considering the role of the mediator variables, PSR and BE on the association of SI→I2P. As it can be seen in Figure 5, the latent variables: EXP, TW, ATT, and SIM, loaded significantly ($\rho$–value≤0.05) on the second–order variable SI, with the following path coefficients: 0.330 with 30.47 *t*-value, 0.361 with 24.09 *t*-value, 0.194 with 11.34 *t*-value, and 0.272 with 22.47 *t*-value, respectively. To test hypotheses *H1*, *H2–H3*, and *H7–H9*, presented in subsections 2.1, 2.2, and 2.4, respectively. The path coefficients and their corresponding $\rho$–values are shown in Figure 5.

---

**Insert Figure 5 about here**

---

Table 12 displays the estimated standardized coefficients and their significance levels. All the hypotheses are accepted with a significant $\rho$–value≤0.05. The *F*–Square values of all constructs are evaluated at the significance level $\rho$–value ≤0.05. The *F*–Square and *R*–Square values are all evaluated above 0.20 and 0.25, respectively, thus, they are significant (Cohen, 1988). The path coefficients of the accepted hypotheses *H1–H3*, and *H7–H9* of the model as:

- $\beta$=0.444 with *F*–Square=0.194 as medium effect for SI→I2P
- $\beta$=0.786 with *F*–Square=0.599 as large effect for SI→PSR
- $\beta$=0.265 with *F*–Square=0.276 as medium effect for PSR→I2P
- $\beta$=0.418 with *F*–Square=0.257 as medium effect for SI→BE
- $\beta$=0.362 with *F*–Square=0.352 as large effect for PSR→BE
- $\beta$=0.184 with *F*–Square=0.297 as medium effect for BE→I2P

The *R*–Square values are all above 0.50 (see Table 12), which suggests a strong statistical power (Cohen, 1988). Regarding the global fit indices, following Hu and Bentler (1998), the SRMR value is evaluated as 0.071, which is below the acceptable 0.080 threshold. The NFI value of our model is evaluated as 0.940, which is below the acceptable 0.90 threshold. Thus, no evidence of model misspecification is detected (Bentler and Bonett, 1980).

We use the *t*-value and its corresponding $\rho$-value as indicators to test the hypotheses, *H1–H3* and *H7–H9* are involved in testing the coefficient of each mediation relationship in the model (see Table 15). So, we assess the mediation role of BIF via PSR on the association

SI➔I2P by first evaluating the mediation role of PSR and BE on SI➔I2P through the following:

- The path coefficient of the indirect effect of SI➔I2P through the mediator variable PSR is found as $\beta$=0.209 with $t$-value= 17.768 and $\rho$–value=0.000
- The path coefficient of the indirect effect of SI➔I2P through the mediator variable BE is found as $\beta$=0.077 with $t$-value= 17.57 and estimated $\rho$–value= 0.009
- The path coefficient of the indirect effect of the sequential mediation of BE via PSR, denoted by SI➔PSR➔BE➔I2P is found as $\beta$= 0.052 with $t$–value=17.510 and the estimated $\rho$–value= 0.012
- The coefficient of the total effect of SI➔I2P is found as $\beta$=0.782 with $t$–value= 28.761 and the estimated $\rho$–value=0.000
- The path coefficient of the direct effect of SI➔I2P is found as $\beta$=0.444 with $t$–value=25.956 and the $\rho$–value= 0.000, which is less than the total effect of SI➔I2P, this is due to the sequential mediating role of PSR with BE.

The total indirect effects amount to 0.338 which results from the sum of all specific indirect effects (0.209+0.077+0.052). These results indicate that the relationship between SI and I2P is partially mediated by PSR and BE, and that the mediation is positive complementary (Demming et al., 2017).

Second, we evaluate the mediation role of PSR on SI➔BE through the following:

- The path coefficient of the indirect effect of the mediator variable PSR on SI➔BE is found as $\beta$=0.285 with $t$–value=18.059 and $\rho$–value=0.000
- The coefficient of the total effect of SI➔BE is found as $\beta$=0.703 with $t$–value=18.36 and $\rho$-value=0.000
- The path coefficient of the direct effect of SI➔BE is found as $\beta$=0.418 with $t$–value=19.962 and $\rho$ -value=0.000, which is less than the corresponding coefficient of the total effect of SI➔BE, this is due to the mediating role of PSR.

These results indicate that the relationship between SI and BE is partially mediated by PSR, and that it is positive complementary (Demming et al., 2017). Third, we evaluate the mediation role of BE on PSR➔I2P through the following:

- The path coefficient of the indirect effect of the mediator variable BE on PSR➔I2P is found as $\beta$= 0.067 with $t$–value=18.570 and $\rho$–value=0.010
- The coefficient of the total effect of PSR➔I2P is found as $\beta$= 0.332 with $t$–value=17.612 and $\rho$–value=0.000
- The path coefficient of the direct effect of PSR➔I2P is found as $\beta$=0.265 with $t$–value = 18.671 and $\rho$–value=0.000, which is less than the total effect of PSR➔I2P, this is due to the mediating role of BE on this association.

These results indicate that BE partially mediates the relationship between PSR and I2P and that it is positive complementary (Demming et al., 2017). In summary, the sum of all coefficients of all specific indirect effects of this model is 0.690 (0.209+0.077+0.052+0.285+0.067). While the sum of all direct effects is evaluated as 1.127 (0.444+0.418+0.265). The sum of direct and indirect effects is equivalent to the sum of all total effects evaluated as 1.817 (0.782+0.703+0.332).

---

**Insert Table 15 about here**

---

We use the open–source package *PLSpredict*, of *SmartPLS* 3.0 to confirm the privilege of the structure of the model in achieving better predictions in comparison to the LM. The RMSE reductions of the PLS model compared to the LM for the observed items of the dependent variable, I2P–1, I2P–2, and I2P–3 are evaluated as 2.48%, 1.14%, and 0.59%, respectively (Table 13). Moreover, the RMSE reductions of the PLS model compared to the LM for the observed items of the independent variables PSR and BE, are all positive, indicating higher error terms of the LM compared to the PLS model (Table 13). Regarding predictive relevance, the $Q$–Square values of the PLS model compared to the LM indicates that the PLS model outperformed the LM. Thus, we confirm the privilege of the PLS structural model in achieving better predictive power.

## 4.6. Sequential Mediation: Prediction Accuracy of the Model Including Benign Envy

We follow the same steps applied in section 5.4 and use the same classification algorithms with 10-fold cross-validation technique to quantify the predictive performance of the sequential mediation model. Table 14 reports the performance metrics of the confusion matrix resulting from the highest–performing algorithm for each path in the structural model.

The prediction accuracy of *H1*: SI➔I2P; *H2*: IS➔PSR; and *H3*: PSR➔I2P are 69.01%, 70.99%, and 68.45%, respectively (see Table 14). For *H7*: PSR➔BE, *H8*: BIF➔I2P, and *H9*:SI➔BE, the accuracy scores are 67.61%, and 68.17%, and 67.04%, respectively (see Table 14). Since all accuracy scores are above 60%, the prediction accuracy for *H1*, *H3*, *H7*, *H8*, and *H9* are considered acceptable while for *H2* is considered good.

All classifiers produced high accuracy scores for all hypotheses except the AdaBoost M1 classifier which performed slightly lower for *H1*, *H2*, and *H8*. The model's prediction accuracy with all associations is 79.73% which is considered good and enhanced compared to 68.17% produced by the model without BE (see Table 14). Therefore, it is recommended to consider the sequential mediation of BE via PSR on the association of SI➔I2P instead of just the mediation effect of PSR.

## 5. Comparative Analysis

In this section, we conduct within–study comparative analysis to assess whether the sequential mediation of PSR with BIF of study – I, and PSR with BE of study – II has any privilege over the mediation effect of just PSR on the association SI→I2P. We also conduct between-studies comparisons for these two sequential mediation models to identify which sequential mediation model provides better predictions of I2P.

In subsection 6.1, we conduct comparative analysis of the mediation effect of the mediator variable PSR with the each of the mediators BIF and BE separately on the association SI→I2P. In subsection 6.2, we compare the predictive accuracy of the models with and without each of BIF and BE separately by assessing the accuracy scores resulting from the confusion matrix of the best-performing ML classifiers. We also assess the descriptive and predictive power of the models with and without the inclusion of each of the variables BIF and BE, separately. In subsection 6.3, we perform further analysis to eliminate speculation of sample bias by applying the split–test ML approach for out–of–sample predictions.

### 5.1. Mediation Analysis

We conduct the mediation comparative analysis by evaluating the difference between the coefficients of the total effect (TE), and the corresponding coefficients of the direct effect (DE) of each model on the association SI→I2P, we call these values as TE and DE in the following:

(i) The difference (TE – DE) of the models with and without the sequential mediation effect in each study independently. We refer to this as within–study comparative mediation analysis.

(ii) The difference (TE – DE) of the sequential mediation models between the two studies, we refer to this as between–studies comparative analysis.

We examine the following mediation effects by evaluating (TE – DE) and also the ratio of the indirect effect (IE) over the TE (Agler and De Boeck, 2017):

(i) SI→PSR→I2P in both studies

(ii) SI→PSR→BIF→I2P in study – I

(iii) SI→PSR→BE→I2P in study – II

(iv) PSR→BE→I2P in study – II

One could verify that PSR→BIF→I2P in study – I is insignificant by checking the $\rho$-value of hypothesis 6 in Table 5, which amounts to 0.510.

**Insert Table 16 about here**

With the intervening effects of the mediators, the TE is partitioned into DE and IE of the mediators. Thus, it is expected for the DE to be lower than the TE in the presence of a significant mediator. The wider the gap between TE and DE, the stronger the role of the mediator on the relationship. The difference (TE – DE) quantifies the indirect effect's contribution to the prediction of the dependent variable, I2P.

For within-study comparisons, we evaluate the proportion of the difference (TE – DE) of the sequential mediation of PSR with BIF over TE as 59.35%, while this proportion for the mediation of just PSR is evaluated as 43.92% (see Table 16, line 4). This indicates that the inclusion of BIF in the model has led to increasing the IE by 15.43% (59.35% – 43.92%), which is greater than 5% and so is significant. Consequently, a higher drop in the DE of the sequential model is observed compared to the mediation of just PSR which amounts to 0.076 (0.226 – 0.150) (see Table 16, line 2).

Similarly, in study – II, we evaluate [(TE – DE)/TE] of the sequential mediation of PSR and BE of study–II as 43.22%, while with the mediation of just PSR as 34.27% (see Table 16, line 4). This indicates that the inclusion of BE in the model has led to increasing the IE by 8.95% (43.22% – 34.27%), which is greater than 5% and so is significant. Consequently, a higher drop in the DE of the sequential model is observed compared to the mediation of just PSR which amounts to 0.068 (0.512 – 0.444), respectively (see Table 16, line 2). These results indicate the privilege of considering the sequential structural models in both studies compared to the single-mediator model.

For further within-study comparisons, we consider the ratio of the IE over TE for the sequential structural models as: (i) SI→PSR→I2P for both studies, (ii) SI→PSR→BIF→I2P and SI→PSR→BE→I2P in study – I and –II, respectively, and (iii) SI→BE→I2P in study – II. One can easily verify that the effect of SI→BIF→I2P of study – I is insignificant (see Table 5). For (i), we evaluate (IE/TE) for SI→PSR→I2P in the model of study – I as 46.88% (0.173/0.369), while in study – II as 26.73% (0.209/0.782) (see Table 16, line 10). For (ii) we evaluate (IE/TE) for SI→PSR→BIF→I2P of study – I as 12.46% (0.046 / 0.369), while for the corresponding SI→PSR→BE→I2P of study – II as 6.64% (0.067 / 0.782) (see Table 16, line 7). For (iii), we also evaluate the (IE/TE) for SI→BE→I2P as 9.85% (0.077/0.782)(see Table 16, line 15).

This indicates that 78.99% (46.88% /59.35%) of the total IE is attributed to PSR in the sequential model of PSR with BIF, compared to 61.84% (26.73%/43.22%) of the total IE attributed to PSR in the sequential model of PSR with BE in study – II. The higher role played by PSR in study – I indicates a lower role played by the second mediator, BIF, which amounts to 21.01% (100% – 78.99%). On the other hand, the second mediator, BE, in study – II contributes 38.16% (100% – 61.84%). This shows a higher contribution of BE compared to BIF in predicting I2P.

## 5.2. Predictive Power and Accuracy

We assess descriptive and predictive power criteria and predictive accuracy of the models in both studies. For descriptive and predictive power, we examine:

(i)    The total variance obtained from the measurement model

(ii)   The model selection criteria, Akaike Information Criterion (AIC), obtained from the measurement model

(iii)  The fit indices of the structural model

(iv)   The RMSE related to the observed items and predicted values obtained from the dependent latent variable for both the PLS model and the simple LM. We compute the percentage of $[(RMSE_{LM} - RMSE_{PLS})/RMSE_{PLS}]$ (see Tables 6 and 13)

(v)    The prediction accuracy of the model resulted by following the classification–based ML algorithms

For within–study comparisons, we use the percentage difference measure, presented in the methodology in subsection 3.2. As shown in Table 17, by the absence of the intervention of BIF, the total variance is decreased by a percentage difference of 6.97, while by the absence of BE, the total variance is decreased by a percentage difference of 13.51. Regarding the information loss, the AIC in the absence of BE and BIF is higher and the percentage difference in the absence of BE is evaluated as 2.33%, while in the absence of BIF is evaluated as 3.29%.

For between-studies comparisons, as shown in Table 17, the percentage increase in the total variance explained is in favor of BE compared to BIF and amounts to 1.96% ((48.512% − 47.578%)/47.578%). It is seen that BE slightly outperforms BIF in reducing the information loss by 0.96% (3.29% − 2.33%). In principle, performing comparisons of AIC values resulting from unidentical samples is invalid (Baguley, 2012, p. 402). The penalty of the model, presented by the AIC measure, is increased by a percentage difference of 3.29 in the absence of BIF, while the penalty in the absence of BE is increased by a percentage difference of 2.33. The presence of BE in the model led to less penalty than BIF (see Table 17).

### Insert Table 17 about here

As shown in Table 17, the model fit indices SRMR and NFI are higher in the models including BIF and BE compared to the corresponding models not including these variables. This indicates the importance of the sequential mediation effect of BIF with PSR and also BE with PSR to fit the data better. In terms of comparing the effect of BIF to that of BE, we note that, in the absence of BE, the increase rate of SRMR and NFI are 5.48% and 2.15%, respectively. These values are higher than that in the absence of BIF which are evaluated as 1.38% and 1.08%, respectively. So, between–studies comparisons indicate that the mediating effect of BE leads to a better model fit.

We also compare the prediction power of the structural models with and without BIF and BE, respectively. This is done by comparing the average of the percentage decrease of RMSE values of the observed items of the dependent variable I2P. The results show that, the percentage difference of RMSE of the structural model including BIF compared to not including it, is reduced by a percentage difference of 7.91 $[(0.92\% - 0.85\%)/((0.92\% + 0.85\%)/2)]$. On the other hand, the absence of BE led to a much higher percentage difference of 114.61 $[(1.40\% - 0.38\%)/((1.40\% + 0.38\%)/2)]$. Moreover, for between-studies comparisons, the inclusion of BE in the model led to a higher average decrease in residuals, which amounts to 1.40%, compared to the effect of the inclusion of BIF, which amounts to 0.92%. This indicates that BE outperforms BIF in predictive power by 52.17%.

Regarding predictive accuracy, the results of the ML classification models indicate that BE also outperforms BIF. The percentage difference between the model's accuracy, including BIF and not including it, is evaluated as 8.47, while the corresponding percentage difference for BE is evaluated as 15.63. The between–studies comparison of the accuracy scores of the model including BIF and including BE yields an increased rate of 13.61% ([79.73% – 70.18%]/70.18%) in favor of BE. This indicates an enhanced predictive accuracy of the model by the sequential mediation of BE with PSR.

## 5.3.    Further Analysis

To eliminate the likelihood that possible variations in the independent samples have a role in steering the results in favor of a specific variable, we applied the split–test ML approach with the coefficients of the PLS model to train the data on one sample and test it on another. The split–test technique is applied by (i) training the variables SI, PSR, and I2P collected in study – I by using PLS–SEM analysis, then (ii) we test the model by applying the trained model to predict the data points collected in study – II for testing. Finally, (iii) we compute the MSE and RMSE of the latent variables I2P and PSR obtained from the training and the testing datasets (see Table 18). Likewise, we use the model trained by the data in study – II to predict the data points of the sample of study – I and examine the error terms.

**Insert Table 18 about here**

We assess the drop or spike in the error terms between the testing dataset compared to the training dataset. Table 18 shows that in the first case, where the data is trained on the sample of study – I and is tested on study – II's sample, the errors in the testing dataset dropped by a percentage difference of 58.32% in MSE and 53.27% in RMSE for I2P and also dropped by a percentage difference of 16.80% in MSE and 7.88% in RMSE for PSR. Thus, the model trained on the data of study – I perform better in predicting the datapoints in the testing dataset.

In the second case, the data is trained on the sample of study – II and is tested on study – I's sample. The error terms resulting from the testing dataset has increased by a percentage difference of 58.17% in MSE and 48.52% in RMSE for I2P and by 17.88% in MSE and 8.63% in RMSE for PSR.

The results of further analysis indicate that the PLS model constructed from the data of study – I yields better out–of–sample predictions in another independent sample. However, our preceding predictive accuracy results are in favor of the sequential PLS model constructed from the data of study – II. Thus, the assessment of the error terms in the testing datasets is not particularly in favor of the dataset of study – II which is concerned of studying BE. Therefore, we could eliminate any possible influence of the sample bias to steer results in favor of BE.

# 6. Conclusion

## 6.1. Discussion of Key Findings

We evaluate the effect of social media influencers' *source influence* SI on the *intention to purchase* I2P by extending the mediation model of *parasocial relationship* PSR on the association of SI → I2P. The extended model considers the sequential mediation of PSR with: (i) *Brand-influencer fit* BIF and (ii) *Benign envy* BE, in two independent studies. We collected data from European and Southeast Asian participants for analysis. The participants were mainly young, aged between 17–27. About 80% of participants of study – I and 100% of the participants in study – II, were active on Instagram for a minimum of three hours weekly.

The *Elaboration Likelihood Model* ELM (Petty and Cacioppo, 1986) has been employed to understand the effectiveness of SMIs' endorsements in relation to audience–related and brand-related features (Xie and Feng, 2022; Yushanlouyi et al., 2022; Masuda et al., 2022). Based on the elaboration likelihood model, we study three main factors associated with the credibility theory and parasocial relationships: (i) the source of the persuasion, SI (ii) the brand–related features, and (iii) audience–related features. Therefore, considering the elaboration likelihood model, we outline the variables of this study as follows:

(i) **Source Influence SI of social media influencers**. We employed the credibility theory (Munnukka et al., 2016) and parasocial theory (Stever, 2017) as in some research studies (Bi and Zhang, 2022; Masuda et al., 2022; Yılmazdoğan et al., 2021) to test the effect of *source influence* SI on the followers' *intention to purchase* I2P. Regarding the constructs defining SI, we observe that both samples indicate a higher role of *trustworthiness* TW and *expertise* EXP in forming SI. TW gives consistent coefficients across the four models of both studies which amounts to about 0.360. The second ranked variable is EXP with coefficient values amounted to 0.320 and 0.330 in study – I and – II, respectively. In study – I, *attractiveness* ATT is the third ranked with a coefficient of 0.253 while *similarity* SIM is third ranked in study – II with a coefficient over 0.260. As for the fourth rank, in study – I, SIM is at the end of the rank with a coefficient of 0.232 in both models and ATT is ranked fourth in study – II with coefficients of about 0.190 (see Figures 2 – 5).

(ii) **Brand–influencer fit BIF as a brand–related feature**. We study the effect of BIF explained by the self–consistency theory (Korman, 1974) to understand why followers are more likely to keep congruence to maintain a balance and preserve their self–image. Predictive power analysis and ML classifiers confirm the validity of the BIF-PSR mediation effect to form practical and managerial implications about the effect of SI on I2P of young Instagram followers.

(iii) **Benign envy as audience or followers–related feature**. We study the effect of benign envy explained by the Self–esteem and self–evaluation theories to hypothesize the sequential mediation effect of PSR and BE (Tran et al., 2022; Lee et al., 2022). Predictive power analysis and ML classifiers confirm the validity of the BIF-PSR mediation effect to form practical and managerial implications about the effect of SI on I2P of young Instagram followers.

Through hypothesis testing of different mediation effects, in subsections 4.3 and 5.3, the significance of the path coefficients of the structural models considering the mediation effect of PSR are confirmed (see Figures 2 and 4). Thus, we conclude the significance of the mediation effect of PSR on the association of SI→I2P. By considering the mediators, BIF and BE with PSR in subsections 4.5 and 5.5, respectively, the significance of the path coefficients of the sequential mediation models are also confirmed (see Figures 3 and 5). Thus, we conclude the significance of the sequential mediation effect of PSR with BIF and PSR with BE on the association of SI→I2P. Despite the extensive application of the parasocial theory in influential marketing, very limited research explored the sequential mediation effects of PSR with other variables (see for example: Hugh et al., 2022, Bi and Zhang, 2022).

Our second contribution is presented by introducing a paradigm for comparative analysis using multidisciplinary methods. This study is the first to incorporate sequential mediation analysis and ML classifier algorithms into a comparative paradigm. Our comparative analysis is conducted on the basis of within–study and between–studies comparisons as follows:

▪ **Within-study comparative analysis**. This compares the mediation effect of just PSR to the sequential mediation effect of PSR with: (i) BIF in study – I, and (ii) BE in study – II. Through mediation analysis, the study shows that the sequential mediation models have higher total indirect effects than the mediation effect with just PSR (see Table 16). This indicates that sequential mediation enhances the performance of the model by including the variables BIF in study – I and BE in study – II. Table 17 shows such enhanced performance of the sequential mediation models through providing higher: (i) total variance explained, (ii) amount of information provided, and (iii) model fit indices, SRMR and NFI.

The results also show a higher predictive power of the sequential mediation of PSR with BIF and PSR with BE compared to the model with just PSR. This is confirmed by assessing the average percentage decrease in the RMSE values resulting from the *PLSpredict* analysis for I2P (See Table 17). To further confirm this result, we apply classification–based ML algorithms to evaluate the prediction accuracy of the models. The results indicate a higher predictive accuracy of the sequential mediation of PSR with BIF and PSR with BE compared to the mediation of just PSR (See Table 17). Thus, with the inclusion of BIF and BE, the models' performance has been enhanced compared to considering just PSR.

- **Between-studies comparative analysis**. This compares the sequential mediation effect of PSR with BIF in study – I with the sequential mediation effect of PSR with BE in study – II. The results shows that the model with BE has a slightly higher indirect effect than BIF (0.667 vs. 0.690). To further confirm this result, we apply classification–based ML algorithms to evaluate the prediction accuracy of the models. The results indicate a higher predictive accuracy of the sequential mediation model of BE with PSR compared to BIF with PSR (70.18% vs. 79.73% See Table 17). Therefore, we recommend considering audience-related features in relation to brand-related features within the framework of credibility theory and parasocial relationships.

To eliminate the possibility that variations in the two samples may have a role in steering the results in favor of a specific variable, we use the split–test ML approach to assess the prediction accuracy of the model with the mediation effect of just PSR, this is due to the availability of the variable PSR in both studies. The model trained on the sample of study – I and tested on the sample of study – II performed better in reducing the RMSE than the model trained on the sample of study – II and tested on the sample obtained from study – I. This result is not particularly in favor of the dataset of study – II, which is concerned with examining the effect of BE. We note that BE was shown to have a privilege over BIF in predicting I2P with higher accuracy score.

Considering these results, we could state with a reasonable confidence that PSR along with BE toward an influencer play a more vital role than PSR with the fit between the influencer and the brand BIF to positively incite the purchase intention of young (17 – 27 years old) followers on Instagram. In this paper, we also show how to use a multidisciplinary approach combining PLS–SEM technique from social science, and machine learning from data analytics to quantify the PLS model's performance. We also propose a comparison paradigm that includes within-study and between-studies analysis.

## 6.2.     Managerial Implications

Managerial and practical implications of SEM should be based on out–of–sample model fit indices (Sarstedt and Danks, 2022). In–sample fit indices are used as indicative of the model's descriptive power (Hair and Sarstedt, 2021). With out–of–sample fit indices, the generalization of practical and managerial implications can be statistically verified and could show its predictive power (Sarstedt and Danks, 2022). *PLSpredict* just compares predictive errors and $Q$-square values in the PLS and simple LM. It does not quantify the model's predictive performance. By applying ML classification algorithms, we obtain some quantifiable accuracy criteria, such as the prediction accuracy score. In support of such consideration, Sarstedt et al. (2022) and Hair and Sarstedt (2021) reviewed the motives of applying machine learning techniques in marketing research.

Stehman (1997) showed that with a prediction accuracy score above 60%, the independent variables predict the dependent variable in a practical context. In our study, all prediction accuracy scores resulting from the best–performing classifiers are found to be above 60%. This is an acceptable threshold for ML–based predictions, and for practical implications.

The findings of this study indicate a better accuracy score in predicting I2P of the mediation model of BE via PSR compared to the mediation model of BIF via PSR, with accuracy scores of 79.73% and 70.18%, respectively. This calls for paying more attention to considering audience–related factors as in BE for studying the I2P in influencer marketing.

The brand-influencer fit is an important factor in the success of the campaign. However, the results suggest that SMIs cast more influence over their audience by inducing benign envy compared to just keeping a brand-fit congruence. Our results also call for considering benign envy in influencer selection strategies than just focusing on the traditional criteria of considering the influencer's large followers' base (Brown and Fiorella, 2013). In summary, instead of putting the SMI or the brand in the center of the influencer selection strategy, the followers–related factors and preferences should be in the heart of the SMI's selection criteria because the followers eventually make the purchase decision (Gross and Wangenheim, 2018).

## 6.3.     Limitations and Further Directions

This study provides several theoretical and practical implications about influencer marketing which faces some limitations. The participants in our study are mainly from SEA and Europe with a limited participation from the U.S., Middle East, Russia, Africa, and other regions where influential marketing is widely practiced. It is recommended to replicate this study in other countries for comparative analysis. This study could benefit from an additional study where both BIF and BE are incorporated into one conceptual framework. Nevertheless, this study motivates examining further sequential and parallel mediation effects in conjunction with the ELM. In addition, it is recommended to investigate how BE and BIF may affect other audience behavioral aspects with an influencer as the *intention to engage* pre– and post – purchase.

## Declaration of Competing Interest

The authors declare that they have no competing financial interests or personal relationships that could have appeared to influence the work reported in this paper.